# MITI Minimum Information guidelines for highly multiplexed tissue images

Denis Schapiro[1,2,3,4,*], Clarence Yapp[1,5,*], Artem Sokolov[1,6,*], Sheila M. Reynolds[7], Yu-An Chen[1], Damir Sudar[8], Yubin Xie[9], Jeremy L. Muhlich[1], Raquel Arias-Camison[1], Sarah Arena[1], Adam J. Taylor[10], Milen Nikolov[10], Madison Tyler[1], Jia-Ren Lin[1], Erik A. Burlingame[11,26], Human Tumor Atlas Network, Young H. Chang[11], Samouil L Farhi[2], Vésteinn Thorsson[7], Nithya Venkatamohan[12], Julia L. Drewes[13], Dana Pe'er[9], David A. Gutman[14], Markus D. Herrmann[15], Nils Gehlenborg[6], Peter Bankhead[16], Joseph T. Roland[17], John M. Herndon[18], Michael P. Snyder[19], Michael Angelo[19], Garry Nolan[19], Jason R. Swedlow[20], Nikolaus Schultz[21], Daniel T. Merrick[22], Sarah A. Mazzilli[23], Ethan Cerami[12], Scott J. Rodig[24], Sandro Santagata[1,24,**] and Peter K. Sorger[1,25,**]

[1] Laboratory of Systems Pharmacology, Ludwig Center for Cancer Research at Harvard, Harvard Medical School, Boston, MA, USA.
[2] Klarman Cell Observatory, Broad Institute of MIT and Harvard, Cambridge, MA, USA.
[3] Institute for Computational Biomedicine, Faculty of Medicine, Heidelberg University Hospital and Heidelberg University, Heidelberg, Germany.
[4] Institute of Pathology, Heidelberg University Hospital, Heidelberg, Germany.
[5] Image and Data Analysis Core, Harvard Medical School, Boston, MA, USA.
[6] Department of Biomedical Informatics, Harvard Medical School, Boston, MA, USA.
[7] Institute for Systems Biology, Seattle, WA, USA
[8] Quantitative Imaging Systems LLC, Portland, OR, USA
[9] Program in Computational and Systems Biology, Memorial Sloan Kettering Cancer Center, New York, NY, USA
[10] Sage Bionetworks, Seattle, WA, USA.
[11] Oregon Health and Science University, Portland, OR, USA.
[12] Dana-Farber Cancer Institute, Boston, MA, USA.
[13] Johns Hopkins University School of Medicine, Baltimore, MD, USA.
[14] School of Medicine, Emory University, Atlanta, GA 30322
[15] Department of Pathology, Massachusetts General Hospital and Harvard Medical School, Boston, MA, USA.
[16] Edinburgh Pathology, Institute of Genetics and Cancer, University of Edinburgh, Edinburgh, UK.
[17] Vanderbilt University School of Medicine, Nashville TN, USA.
[18] Department of Surgery, Washington University School of Medicine, St. Louis, MO USA.
[19] School of Medicine, Stanford University, Stanford, CA, USA.
[20] Division of Computational Biology and Centre for Gene Regulation and Expression, University of Dundee, Dundee, UK.
[21] Department of Epidemiology & Biostatistics at Memorial Sloan Kettering Cancer Center, New York, NY, USA
[22] Pathology, University of Colorado, Aurora, CO, USA.
[23] Boston University, MA, USA.
[24] Department of Pathology, Brigham and Women's Hospital, Boston, MA, USA
[25] Department of Systems Biology, Harvard Medical School, Boston, MA
[26] Current Affiliation: Indica Labs, Albuquerque, NM, USA

\* These authors contributed equally
\*\* Co-Corresponding Authors

Corresponding Authors, Lead Contact
Sandro Santagata, ssantagata@bics.bwh.harvard.edu
Peter Sorger, peter_sorger@hms.harvard.edu

**The imminent release of tissue atlases combining multi-channel microscopy with single cell sequencing and other omics data from normal and diseased specimens creates an urgent need for data and metadata standards that guide data deposition, curation and release. We describe a Minimum Information about highly multiplexed Tissue Imaging (MITI) standard that applies best practices developed for genomics and other microscopy data to highly multiplexed tissue images and traditional histology.**

Highly multiplexed tissue imaging using any of a variety of optical and mass-spectrometry based methods (Supplemental **Table 1**) combines deep molecular insight into the biology of single cells with spatial information traditionally acquired using histological methods, such as hematoxylin and eosin (H&E) staining and immunohistochemistry (IHC)[1]. As currently practiced, multiplexed tissue imaging of proteins involves 20-60 channels of 2D data, with each channel corresponding to a different antibody or colorimetric stain (**Figure 1**). Multiple inter-institutional and international projects, such as the Human Tumor Atlas Network (HTAN)[2], the Human BioMolecular Atlas Program (HuBMAP)[3], and the LifeTime Initiative[4] aim to combine such highly multiplexed tissue images with single cell sequencing and other types of omics data to create publicly accessible "atlases" of normal and diseased tissues. Easy public access to primary and derived data is an explicit goal of these atlases and is expected to encompass native-resolution images, segmented single-cell data, anonymized clinical metadata and treatment history (for human specimens), genetic information (particularly for animal models), and specification of the protocols used to acquire and process the data. Given the imminent release of the first atlases, an urgent need exists for data and metadata standards consistent with emerging FAIR (Findable, Accessible, Interoperable, and Reusable) standards[5]. In this commentary, we establish the MITI (Minimum Information about highly multiplexed Tissue Imaging) standard and associated data level definitions; we also discuss the relationship of MITI to existing standards, practical implementations, and future developments.

**Scope and target audiences**

MITI covers biospecimen, reagent, data acquisition and data analysis metadata, as well as data levels for imaging with antibodies, aptamers, peptides, dyes and similar detection reagents (**Supplementary Table 1**). The standard is also compatible with images based on H&E staining, low-plex immunofluorescence (IF) and IHC. A working group is currently extending MITI to cover subcellular resolution imaging of nucleic acids using methods such as MERFISH[6]. While conceived with today's two-dimensional (2D) images in mind (these typically involve 5 - 10 μm thick sections of fixed or frozen specimens), MITI accommodates three-dimensional (3D) datasets acquired using confocal, deconvolution and light sheet microscopes[7]. MITI has been established as its own organization with its own GitHub repository, governing structure, and procedures for proposing and incorporating revisions. The definition of MITI is available in machine readable YAML format (https://github.com/miti-consortium/MITI) along with other relevant information. MITI has also been implemented in practice (https://github.com/ncihtan/data-models) and used to structure metadata available via the HTAN data portal (https://htan-portal-nextjs.vercel.app). However, MITI is independent of HTAN or any single research consortium.

Highly multiplexed imaging is derived from methods such as IHC and IF that are in widespread use in pre-clinical research using cultured cells and model organisms, and in clinical practice with human tissue specimens. Many standards and best practices have been established for these types of data (**Supplementary Table 2**), but high-plex imaging presents unique challenges: images are expensive to collect and can be very large (up to 1TB in size), specimens are often difficult to acquire and may have data use restrictions, and accurate clinical and genomic annotation is a necessity. Recent interest in highly multiplexed tissue imaging has been driven by applications in oncology, largely due to the importance of the tumor microenvironment in immuno-editing and responsiveness to immunotherapy, but the approach is broadly applicable to studying normal development, infectious disease, immunology and other topics. HuBMAP[3], for example, is using high-plex imaging to study a range of normal human tissues. MITI is also relevant to studies with model organisms and data tables have already been created to store data from genetically engineered mouse models (GEMMs) in a standardized manner.

Multiplexed imaging also promises to impact the pathological diagnosis of diseases, which is rapidly switching to digital approaches[8]. For over a century, histological analysis of anatomic specimens (from biopsies and surgical resection) has been the primary method of diagnosing diseases such as cancer[9], and this remains true today, despite the impact of gene sequencing. Multiplexed tissue imaging promises to augment conventional pathological diagnosis with the detailed molecular information needed to specify use of contemporary precision therapies. This is therefore an opportune time to seek alignment of research and diagnostic approaches by establishing public standards able to take full advantage of the detailed molecular information revealed by emerging imaging methods.

**Existing standards and approaches**

The Human Genome Project, the Cancer Genome Atlas (TCGA)[10] and similar large-scale genomic programs have developed several approaches to data management of immediate relevance to tissue atlases. The first is the concept of "minimum information" metadata, which has been employed in microarrays (the MIAME standard)[11], genome sequences (MIGS)[12], and biological investigation in general (MIBBI)[13]. The second is the idea of "data levels" (https://gdc.cancer.gov/resources-tcga-users/tcga-code-tables/data-levels), which specify the extent of data processing (raw, normalized, aggregated or region of interest, corresponding to data levels 1-4) and access control. Access control is required because even anonymized DNA sequencing data pose a re-identification risk[14]. As a result, the database of Genotypes and Phenotypes (dbGaP), the NCI Genome Data Commons (GDC)[15], and the US Federal Register (79 FR 51345) control access to primary sequencing data (so-called level 1 & 2 sequencing data) based on policies set by a data access committee. Higher level genomic data, which are generally more consolidated, involve information aggregated form many patients, and pose little or no re-identification risk can be freely shared[16] (**Figure 2**). When datasets are combined, they acquire the most stringent restriction applied to any constituent element. While we are not aware of any policies addressing the anonymity of histological images, consultation with our Institutional Review Boards (IRBs, ethics committees) has led us to conclude that public release of tissue images does not constitute a risk to patient privacy. MITI data levels are nonetheless consistent with the existing GDC and dbGaP practice that data intended for unrestricted distribution are classified as level 3 and up. In the case of images adhering to the MITI standard, level 3 data have been subjected to quality control and some degree of human annotation, making them more useful in a shared environment than raw images. We anticipate that IRBs and government agencies will in the future provide further guidance on sharing of datasets that combine clinical history, sequence information, and tissue images; MITI will be adapted to accommodate such guidance.

The MITI standard also draws extensively on image formats developed for cultured cells and model organisms and on a wide variety of open-source software tools (**Supplementary Table 3**). Noteworthy among these are the Open Microscopy Environment (OME) TIFF standard[17] and the BioFormats[18] approach to standardization of microscopy data. MITI field definitions are harmonized with the QUality Assessment and REProducibility for Instruments and Images in Light Microscopy (QUAREP-LiMi)[19] effort, the Resource Identification Initiative[20], and antibody standardization efforts by the Human Protein Atlas[21] and are also compliant with the recently developed Recommended Metadata for Biological Images initiative[22]. Metadata on model organisms (particularly GEMMs - and patient derived xenografts - PDXs) are aligned with existing standards, many developed for genomic information (see **Supplementary Table 2** for a full list of antecedent resources). Well-curated clinical information is essential for the interpretation of data from human specimens but standardizing such information has proven to be a major challenge in the past, for example in TCGA[23]. Thus, HTAN and other current NCI projects focused on human specimens are emphasizing standardization of clinical metadata, and the MITI standard is designed to closely align with the Genomic Data Commons (GDC) Data Model[24] in this regard (**Supplementary Tables 5-6**).

All imaging methods generate data that comprises a sequence of intensity values on a raster; multi-spectral imaging simply adds new dimensions to the raster. The cameras that collect H&E and IHC images from bright-field microscopes or high-plex images from fluorescence microscopes generate a raster; ablation-based mass-spectrometry imaging (e.g. MIBI and IMC) is also raster based. As currently defined, MITI specifies that raster images should be stored in the OME-TIFF 6 standard, but OME formats are currently being migrated to a set of next generation file formats (collectively OME-NGFF)[25] to improve scalability and performance on the cloud. MITI will be updated to align with these new formats as they come into general use. Another area of translational and clinical research in which imaging is commonly encountered is radiology, which is almost entirely digital, and uses data interchange standards governed by DICOM (https://www.dicomstandard.org/). DICOM has recently been extended to accommodate both radiology data and OME-TIFF standards[26]. The NCI's ongoing program to create an Imaging Data Commons[27] is expected to be based on this dual standard, or on a successor using OME-NGFF. MITI is, or will be, compatible with these foundational data standards.

In highly multiplexed tissue imaging antibodies are either conjugated to fluorophores directly or via oligonucleotides, or are bound to secondary antibodies (Figure 1, **Supplementary Table 4**). Images are then acquired serially, one to six channels at a time, to assemble data from 20-60 antibodies. In ablation-based methods, antibodies are labelled with metals and vaporized with lasers or ion beams after which they are detected by atomic mass spectrometry (**Supplementary Table 4)**. In all cases, the raw output of data acquisition instruments comprises Level 1 MITI data (**Figure 2**), analogous to the Level 1 FASTq files in genomics.

Whole slide imaging is required for clinical applications[28] and also necessary to ensure adequate power in pre-clinical studies[29]. However, resolution and field of view have a reciprocal relationship – both with respect to optical physics and the practical process of mapping image fields onto the fixed raster of a camera (or ablating beam). Whole slide images of histological specimens[8] must therefore be acquired by dividing a large specimen into contiguous tiles. This usually involves acquisition of ~100 to 1,000 tiles by moving the microscope stage in both X and Y, with each tile being a multi-dimensional, subcellular resolution, TIFF image. Tiles are combined at sub-pixel accuracy into a mosaic image in a process known as stitching. When high-plex images are assembled from multiple rounds of lower-plex imaging, it is also necessary to register channels to each other across imaging cycles and to correct for any unevenness in illumination (so-called flat fielding)[30]. Stitched and registered mosaics can be as large as 50,000 x 50,000 pixels x 100 channels and require ~500 GB of disk space. They correspond to Level 2 MITI data and represent full-resolution primary images that have undergone automated stitching, registration, illumination correction, background subtraction, intensity normalization and have been stored in a standardized OME format. The level of processing is analogous to BAM files, a common type of Level 2 data in genomics.

Level 3 data represent images that have been processed with some interpretive intent, which may include (i) full-resolution images following quality control or artifact removal, (ii) segmentation masks computed from such images, (iii) machine-generated spatial models, and (iv) images with human or machine-generated annotations. Level 3 MITI data is roughly analogous to Level 3 mRNA expression data in genomics. However, whereas many users of genomic data only require access to processed level 3 and 4 data, which are usually quite compact, quantitative analysis of tissue images adds a requirement for full-resolution primary images so that images and computed features can be examined in parallel[31]. Level 3 MITI data is intended to be the primary type of image data distributed by tissue atlases and similar projects.

Assembled level 3 images are typically segmented to identify single cells[31], which are quantified to produce a "spatial feature table" that describes marker intensities, cell coordinates and other single-cell features. The Level 4 data in spatial feature tables are a natural complement to count tables in single cell sequencing data (e.g. scRNA-seq, scATAC-seq, scDNA-seq) and can be analyzed using many of the same dimensionality reduction methods (e.g. PCA, t-SNE and U-MAP)[32] and on-line browsers such as cellxgene (**Supplementary Table 3**)[33]. These types of tabular data are all examples of "Feature Observation Matrixes" which are themselves being standardized across domains of biology to improve their utility and inter-compatibility. Level 5 MITI data comprise results computed from spatial feature tables or primary images. Because access to TB-size full-resolution image data is impractically burdensome when reading a manuscript or browsing a large dataset, a specialized type of Level 5 image data has been developed to enable panning and zooming across images using a standard web browser. In the case of Level 5 images viewed with MINERVA software, the aim is to exploit similar functionality and concepts as those in Google Maps or electronic museum guides[34]. The inclusion of digital docents with images makes it possible to combine pan and zoom with guided narratives that greatly facilitate comprehension of complex datasets and promote new hypothesis generation[35].

For any metadata standard to be used, a balance must be struck between ease of data entry, which minimizes non-compliance by data generators, and level of detail, which must be sufficient for data retrieval, analysis, and publication in a reproducible manner. Moreover, specifying a metadata standard is separate from the essential task of developing a practical and reliable means for capturing information needed to ensure adherence to the standard. Two approaches have proven most effective in addressing this requirement. One, exemplified by OMeta[36], involves a relational database and web interface that data generators use to input necessary information in a controlled manner. Another approach, exemplified by MAGE-TAB[37], involves a standardized format for collecting metadata via a series of structured documents, which are then used to populate web pages and databases[38]. As a practical test of MITI we have implemented the latter approach in a JSON schema (https://github.com/ncihtan/data-models) that also conforms to the design principles of SCHEMA.org. These principles focus on the creation, maintenance and promotion of schemas for structured data that is supported by major web search engines, thereby enhancing discoverability. In this TAB-like approach the MITI standard is exposed to data collectors as Google Sheets with dropdowns representing controlled vocabularies and highlighting required or optional elements; many fields are automatically validated upon entry. These documents are ingested using SCHEMATIC (Schema Engine for Manifest Ingress and Curation; https://github.com/Sage-Bionetworks/schematic), automatically linked to primary imaging data, and stored as cloud assets. These implementations continue to evolve, and entirely different approaches are possible: nothing in a MITI-type standard constrains how data are collected.

Whereas many research agencies and countries have made a major investment in curating, storing, and distributing genomic data, fewer repositories exist for primary image data. The Image Data Resource[39] maintained by the European Bioinformatics Institute (EBI) is an exception, but as the volume of image data grows, other means of data distribution will almost certainly be required. In the U.S., in the absence of a major public investment in data storage, the development of "requester pays"[40] access to datasets is a promising development. The primary cost associated with creation and maintenance of a dataset on a commercial cloud service involves data download, not data ingress and storage. In a "requester pays" model, a user seeking access to a dataset pays the cost of data egress directly to the cloud provider making access both secure and anonymous (moreover, the cost of egress into another account on the same commercial cloud is low). Although the "requester pays" approach might appear to create an impediment to research, the actual cost of egress is quite low (currently, about hundred US dollars per TB) compared to any form of data acquisition and a key goal is to avoid a tragedy of the commons in which frequent, duplicate downloads overwhelm the system. A combination of a MITI implementation on a cloud service (as described above) with "requester pays" cloud access will also make it possible for individuals to distribute very large FAIR image datasets at relatively low cost. Such an approach does not obviate the need for public investments, such as those being made but EBI, but does represent a practical way forward to democratize release of standardized data – some of which can then be incorporated into publicly supported resources. Regardless, the MITI standard described here is available for immediate use, without being impacted by how access to the primary data is provisioned.

Public data and metadata standards have been essential for the success of genomics and other fields of biomedicine, but the creation of a new standard is no guarantee of successful adoption. An outpouring of effort 10-20 years ago led to the development of widely adopted and well maintained standards such as MIAME[11], MIGS[12] and MIBBI[13], and these have been consolidated and further documented by the Digital Curation Center (https://www.dcc.ac.uk/), FairSharing.org, and similar projects. However many other minimum information projects have been left unattended[41], and it remains unclear whether existing metadata adequately conform to user needs[42]. The development of MITI and of the initial HTAN implementation enjoys NCI support and is expected to become part of the NCI Cancer Research Data Commons[27], helping ensure its viability. However, individuals and organizations are invited to join in the further development of MITI and should make contact via the image.sc forum or submit pull requests (i.e. requests for inclusion in the MITI "code base" at https://github.com/miti-consortium/MITI). Because high high-plex tissue imaging is in its infancy and MITI has attracted the great majority of developers of existing high-plex tissue image acquisition methods, it represents a solid beginning for what will need to be an evolving standard. By having its own repository and governance structure, independent of any particular research program or constituency, MITI also conforms with other requirements of successful open standards[43].

Data and Code Availability Statement

The detailed specification of the guidelines outlined in this manuscript are available at https://github.com/miti-consortium/MITI and https://www.miti-consortium.org/

Acknowledgements

This work is supported by the HTAN Consortium and the Cancer Systems Biology Consortium (CSBC). A list of all current Consortium members can be found at https://humantumoratlas.org/.

This work was supported by the following grants from the National Cancer Institute under the Human Tumor Atlas Network (HTAN) U2C CA233262 (Harvard Medical School), U2C CA233280 (OHSU), U2C CA233195 (Boston DFCI Broad), U2C CA233291 (Vanderbilt University Medical Center), U2C CA233311 (Stanford University), U2C CA233238 (Boston University Medical Campus), U2C CA233285 (Children's Hospital of Philadelphia), U2C CA233303 (Washington University St. Louis), U2C CA233280 (Oregon Health and Science University), U2C CA233284 (Memorial Sloan Kettering Cancer Center), U2C CA233254 (Duke University Medical Center) and by other public support including U54 CA225088 (SS, PKS) and U24 CA233243 (Dana-Farber Cancer Institute, Emory University, Institute for Systems Biology, Memorial Sloan Kettering Cancer Center, Sage Bionetworks). DS was funded by an Early Postdoc Mobility fellowship (no. P2ZHP3_181475) from the Swiss National Science Foundation and was a Damon Runyon Fellow supported by the Damon Runyon Cancer Research Foundation (DRQ-03-20); DS is currently supported by the BMBF (01ZZ2004). NG was funded by the NIH Human BioMolecular Atlas Program (HuBMAP) OT2 OD026677 and MDH by NCI/NIH Task Order No. HHSN26110071 under Contract No. HHSN2612015000031.

Author contributions

D.S., C.Y and A.S. initiated and implemented the MITI guidelines with extensive guidance from other authors and direct supervision by P.K.S. and S.S.. All authors contributed to and reviewed the final MITI guidelines. D.S., C.Y, A.S., P.K.S and S.S. wrote the manuscript with input from all authors.

Competing Interests Statement

PKS is a member of the SAB or BOD of Applied Biomath, RareCyte Inc., and Glencoe Software, which distributes a commercial version of the OMERO data management platform; PKS is also a member of the NanoString SAB and a consultant to Merck and Montai Health. In the last five years the Sorger lab has received research funding from Novartis and Merck. Sorger declares that none of these relationships have influenced the content of this manuscript. SS is a consultant for RareCyte Inc. NG is a co-founder and equity owner of Datavisyn. DS is a consultant for Roche Glycart AG. JRS is Founder and CEO of Glencoe Software, which distributes a commercial version of the OMERO data management platform. SR receives research funding from Bristol-Myers-Squibb, Merck, Affimed, and Kite/Gilead. SR is on the Scientific Advisory Board for Immunitas Therapeutics. DSu is employed by Quantitative Imaging Systems LLC. EAB is an employee of Indica Labs.

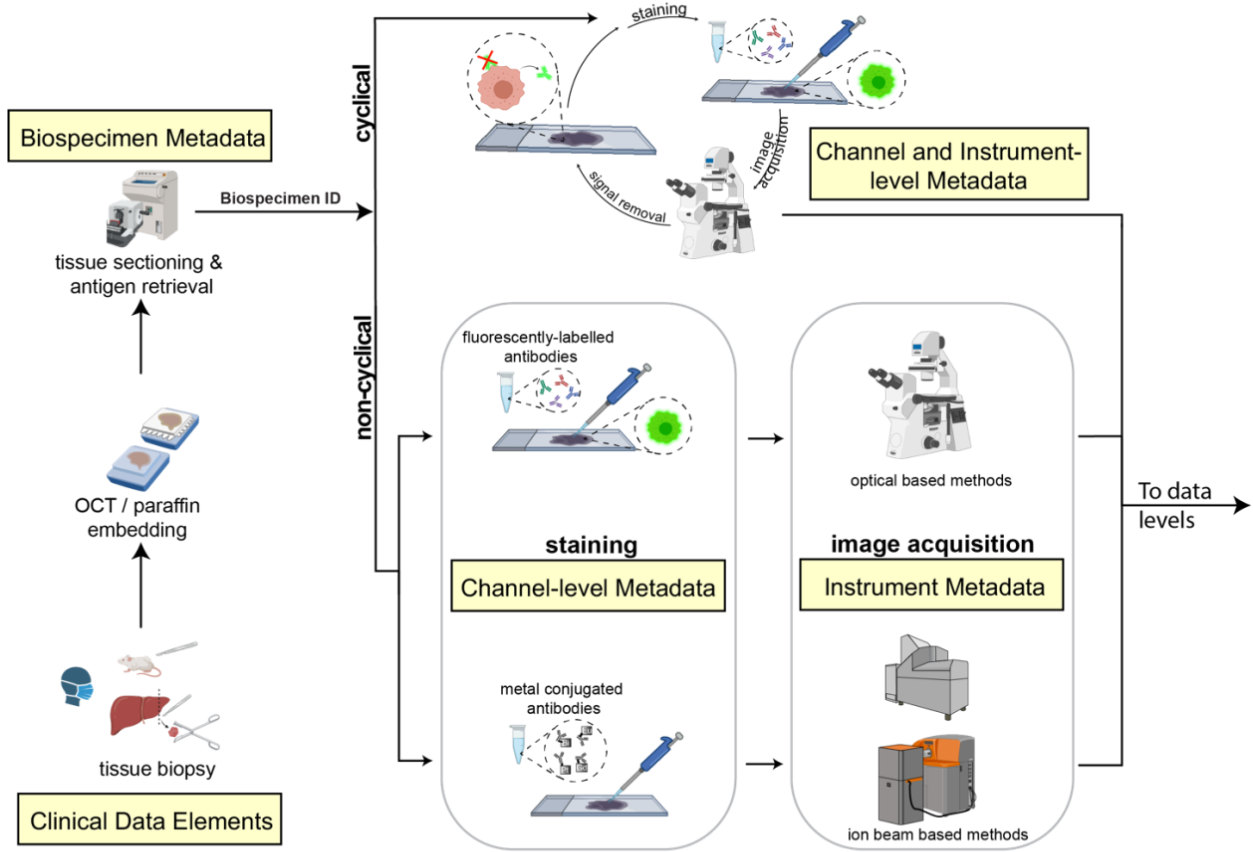

**FIGURE 1: Schematic diagram of the steps in a canonical multiplexed tissue imaging experiment and the associated metadata**

In a typical workflow, samples collected from patient biopsies and resections or from animal models are formaldehyde fixed and paraffin embedded (FFPE) or frozen and then sectioned and mounted onto either a standard glass microscope slide (for CyCIF, mIHC, IMC, MELC or mxIF), fluidic chamber (for CODEX) or specialized carriers (for MIBI). Clinical and biospecimen metadata (extracted from clinical records, for example) is linked to all other levels of metadata via a unique ID (Biospecimen ID). Data is acquired using cyclical or non-cyclical staining and imaging methods and both reagent and experimental metadata collected (consisting of antibody, reagent and instrument metadata). In both cyclic and non-cyclic methods, sections undergo pre-processing, antigen retrieval, and antibody incubation and images are acquired. In cyclical imaging methods, fluorophores or chromogens are inactivated or removed and additional antibodies and/or visualization reagents are applied and data acquisition repeated. Channel and instrument metadata capture these essential details. Created with BioRender.com.

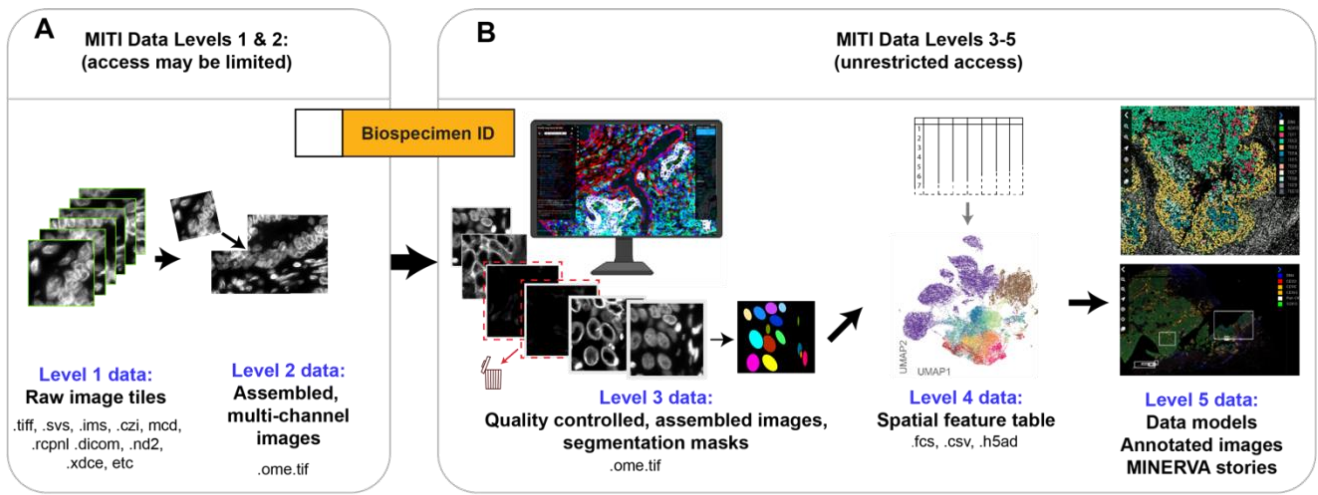

**FIGURE 2: MITI data levels and formats**

Data levels specify the extent of data processing and, in the case of sequencing data, whether access requires the approval of a data access committee. In common practice, data at levels 3 and up are freely shared. Primary data arising from microscopes and data acquisition instruments corresponds to level 1 data. Because the raw image data acquired from one slide usually consists of separate image fields, possibly from proprietary formats, they are processed to correct for uneven illumination and other instrumentation artifacts and assembled into a single multi-channel image in the OME-TIFF format (level 2 data). OME-TIFF image mosaics undergo quality control (including artefact removal, channel rejection, evaluation of staining quality) to generate full-resolution, assembled and curated level 3 image data; segmentation algorithms generate one or more label masks that also comprise level 3 data. The great majority of users will want to access these level 3 images. Each label mask (e.g., nuclei, cytoplasmic-regions, whole cells, organelles, etc.) is used to compute quantitative features, such as the mean signal intensity, spatial coordinates of individual cells and morphological features, which are stored as level 4 spatial feature tables (where rows represent single cells and columns the extracted cellular features); these data are suitable for analysis using the dimensionality reduction and visualization tools used for other types of single-cell data (e.g. UMAP plots). Spatial models computed from images and spatial feature tables, or by direct application of machine learning to images, as well as images annotated by humans, comprise level 5 data.

# Supplementary Materials

**Supplementary Table 1. Description of the components for reporting a highly multiplexed tissue imaging experiment**

| Data Level | Assay data | Assay data description | Metadata | Metadata Description |
|---|---|---|---|---|
| N/A | N/A | N/A | Clinical metadata | Participant identifiers, demographics, diagnosis, exposures, treatments, and follow-up (**Supplementary Table 5**) |
| N/A | N/A | N/A | Biospecimen metadata | Identifying and descriptive information about the method of tissue acquisition, sample processing and handling, and features pertaining to histologic assessment (**Supplementary Table 7**) |
| 1 | Raw numerical output of acquisition instruments | Vendor-specific image formats, ideally Bioformats compatible. Whole slide images recorded as individual tiles from x, y stage positions | File level metadata | Microscope specifications, image acquisition parameters, sample information (**Supplementary Table 8**) |
| 2 | Full resolution OME-TIFF | Primary images after stitching, registration, illumination correction, background subtraction, intensity normalization using automated software. File conversion from vendor specific format to universal OME-TIFF | OME-TIFF header metadata | Image spatial and bit depth properties, channel information (**Supplementary Tables 8-10**) |
| | | | Channel level metadata | Microscope specifications, channel information, fluorophore/metal label data, experiment protocol (**Supplementary Tables 8-10**) |
| 3 | Full resolution OME-TIFF | Level 2 image data after quality control to remove image artefacts and uninformative channels caused by tissue degradation or poor antibody staining | OME-TIFF header metadata | Image spatial and bit depth properties, channel information, rendering settings for visualization e.g., Minerva, OMERO, etc. |

| | | | | (**Supplementary Tables 8-10**) |
|---|---|---|---|---|
| | | | Channel level metadata | Microscope specifications, channel information, fluorophore/metal label data, experiment protocol (**Supplementary Tables 8-10**) |
| | Segmentation masks as OME-TIFF | Labelled masks where each cell has a unique ID. Masks generated by automated segmentation models and with some human oversight | Segmentation level metadata | Image spatial and bit depth properties, algorithm/workflow/model details (**Supplementary Table 11**) |
| 4 | Spatial feature tables | Single cell features such as marker intensities, centroid coordinates, and area. Derived from level 2 and 3 | Spatial feature table metadata | Image spatial and bit depth properties, algorithm/model details (**Supplementary Table 12**) |
| 5 | Data exploration | Cell type annotations | Cell state level metadata | Cell type annotations, image derived data, channel information, algorithm/workflow/method (ie. UMAP, t-SNE), rendering settings for visualization (Minerva, OMERO, etc) (**Supplementary Tables 12, 13**) |
| | | Dimensionality-reduction from spatial feature table | | |
| | | Pixel-level machine learning | | |
| | | Annotated visualization and data exploration | | |

An example of a structured data base for these components is available at https://htan-portal-nextjs.now.sh/explore. In addition, documentation for automated data ingress, as implemented in the HTAN consortium, is available at https://ncihtan.github.io/HTAN-Data-Ingress-Docs/.

**Supplementary Table 2. Overview of relevant other initiatives / standards which guided MITI reporting guidelines**

| Initiative / Standard |
|---|
| "Standardized" validation of antibodies for research applications. Human Protein Atlas[1] |
| The Resource Identification Initiative[2] |
| A Global View of Standards for Open Image Data Formats and Repositories[3] |
| QUAREP-LiMi: A community-driven initiative to establish guidelines for quality assessment and reproducibility for instruments and images in light microscopy[4] |
| Minimum Information guidelines for fluorescence microscopy: increasing the value, quality, and fidelity of image data[5] |
| Minimum Information About a Microarray Experiment (MIAME)[6] |
| Minimum Information about a Genome Sequence[7] |
| Guidelines for reporting single-cell RNA-seq experiments - minSCe[8] |
| Minimum Information for Biological and Biomedical Investigations[9] 23/02/2022 17:04:00 |
| Minimum Information Specification For In Situ Hybridization and Immunohistochemistry Experiments (MISFISHIE)[10] |
| MIRO: guidelines for minimum information for the reporting of an ontology[11] |
| BINA: 4D Bioimaging North America (BINA)[12] |
| MethodsJ2[13] |
| Sharing biological data: why, when, and how[14] |
| PDX-MI: Minimal Information for Patient-Derived Tumor Xenograft Models[15] |
| REMBI: Recommended Metadata for Biological Images—enabling reuse of microscopy data in biology[16] |

**Supplementary Table 3. Open source visualization and analysis tools tailored for highly multiplexed imaging methods**

| Software | Description | Implementation | Source/ Link |
|---|---|---|---|
| napari[17] | Image viewer | local | https://github.com/napari/napari |
| Facetto[18] | Image viewer and image analysis | cloud-based | https://github.com/kruegert/facetto |
| histoCAT[19] | Image viewer and image analysis | local and cloud-based | https://github.com/BodenmillerGroup/histoCAT |
| QuPath[20] | Image viewer and image analysis | local | https://qupath.github.io/ |
| Minerva[21] | Data sharing, image viewer and image analysis | local and cloud-based | https://github.com/labsyspharm/minerva-story |
| MCMICRO[22] | Image preprocessing and image analysis | local and cloud-based | https://github.com/labsyspharm/mcmicro |
| Viv[23] | Image visualization library | cloud-based | https://github.com/hms-dbmi/viv |
| ImaCytE[24] | Image viewer and image analysis | local | https://github.com/biovault/ImaCytE |
| OMERO[25] | Image data management and viewer | cloud-based | https://github.com/ome/openmicroscopy |
| ImageJ[26] | Image viewer and image analysis | local | https://imagej.net |
| Giotto[27] | Image viewer and image analysis | local | https://github.com/RubD/Giotto/ |
| cellxgene[28] | Interactive explorer for single-cell transcriptomics data | local and cloud-based | https://github.com/chanzuckerberg/cellxgene |
| Squidpy[29] | Image viewer and image analysis | local | https://github.com/theislab/squidpy |

**Supplementary Table 4. Highly multiplexed imaging methods**

| Optical | Fluorophore-based | Single staining and cyclic detection | CO-Detection by indEXing (CODEX)[30] |
| --- | --- | --- | --- |
| | | | Immunostaining with Signal Amplification By Exchange Reaction (Immuno-SABER)[31] |
| | | Cyclic staining and detection | Cyclic Immunofluorescence (CyCIF)[32] |
| | | | Multiplex immunofluorescence (MxIF)[33] |
| | | | Iterative indirect immunofluorescence imaging (4i)[34] |
| | | | Multi-epitope-ligand cartography (MELC)[35] |
| | Enzyme-based | | Multiplex Immunohistochemistry (mIHC)[36] |
| | | | Multiplexed immunohistochemical consecutive staining on single slide (MICSSS)[37] |
| Non-optical | Detection based on atomic mass spectrometry of metal-labelled antibodies | | Imaging Mass Cytometry (IMC)[38] |
| | | | Multiplexed Ion Beam Imaging (MIBI)[39] |

**Supplementary Table 4:** An overview of the most commonly used antibody-based highly multiplexed tissue imaging methods. Low-plex imaging methods, such as Hematoxylin and Eosin (H&E), Immunohistochemistry (IHC), Vectra and others are supported but not included in the table.

# Supplementary Note

## 1. The components needed to describe a highly multiplexed tissue imaging experiment

The proposed metadata schema is currently designed to record information from tumor atlas samples and includes clinical data elements to provide information about biospecimens from human subjects with either cancer or precancer lesions. However, the data elements presented here can be adapted for experiments utilizing samples from patients with a range of disease types and the framework can be modified to accommodate non-human samples as well as non-tumor samples. A minimum set of clinical data elements is provided here (**Supplementary Table 1**) that covers participant demographics, diagnosis, exposures, molecular testing, treatments and follow-up. In the HTAN initiative (https://humantumoratlas.org), attributes are available for reporting additional characteristics such as prior patient history, family history, and molecular testing. The complete HTAN data standards are browsable at https://htan-portal-nextjs.now.sh/standards.

Following the example of guidelines for reporting genomic data, we have indicated the level of significance for these attributes as either "required" or "recommended." A value (specified below) must be added for attributes marked as required. The values will be represented by strings, dates, numeric or boolean variables, as appropriate. Whenever possible, valid values will be limited to either a predefined set of keywords (e.g., "Alive" or "Dead") or a numeric interval (e.g., positive integer). Values of 'Unknown', 'Not Reported' and 'Not Applicable' can be used if information is missing for certain samples. 'Unknown' indicates that it is unclear whether data was or was not reported, 'Not Reported' indicates that the information was never collected and cannot be updated, and 'Not Applicable' indicates that the attribute does not apply to the participant or the study. When "required" attributes are conditional (e.g., only available for a specific method/technology or only relevant to a specific disease type) 'Not Applicable' should be used. 'Unknown' or 'Not Reported' should be used whenever possible. The number '0' represents the numerical value.

Submitters of image data must remove Protected Health Information (PHI) and must de-identify the data prior to submission. Participant PHI that must be removed according to the safe harbor method for de-identification includes names, medical record numbers, geographical identifiers smaller than a state, contact information, all dates (other than year) related to the participant (including birthdate, encounter dates such as procedure and follow up dates, date of death). One example of an automated de-identification procedure is provided by DICOM, and some repositories may require stringent removal of all absolute dates even if they are not directly related to PHI such as the date of image acquisition.

To obfuscate PHI, dates are reported as calculated fields (in days) using reference or anchor dates ("index date") as described by the GDC. Recommended index dates are included in the attribute descriptions (**Supplementary Tables 5, 6**). As an example, the HTAN initiative utilizes the date of birth of the participant as the index date. TCGA used the date of pathologic diagnosis as the index date.

To enable effective processing by computational tools, all metadata files must follow tidy data practices[40]. Specifically, metadata tables should have rectangular shape with rows corresponding to observation instances (e.g., patients, channels, markers, etc.) and columns corresponding to attributes **(Supplementary Tables 5-12)** measured in each instance. Adherence to tidy data standards is straightforward when metadata tables are stored using standard formats like comma-separated value (CSV) or H5AD[41]. When composing metadata tables manually, users are cautioned that Microsoft Excel is known to corrupt data upon entry[42]; the issue can be avoided by using alternative spreadsheet software, e.g., Google Sheets.

**Implementation and maintenance of the MITI standard in YAML**
The metadata definitions in this manuscript (Supplementary Table 1) represent a conceptual design for a standard in human-readable form. General definitions are accompanied by detailed specifications, also in the form of human readable tables, for metadata fields and their allowed values (Supplementary Tables 5-12). To enable implementation of MITI in a real-world setting, the specification is available in a machine-readable YAML format through a publicly accessible GitHub repository (https://github.com/labsyspharm/MITI). It is the intention of MITI developers to build software tools to implement MITI based on these YAML files.

The choice of YAML as a markup language was motivated by its simple declarative structure making it both machine and human readable. Individual YAML files capture attributes, their description and significance (Supplementary Tables 5-12), but also additional information that is essential for validating specific files; this includes ensuring that data in each field has the correct attribute (e.g., boolean, integer, string, etc.) and that it meets constraints on valid values (e.g., a predefined set of keywords). MITIv1.0 YAML files are an exact match to the content presented here, but MITI is expected to undergo regular updates as it is deployed on a large scale by the imaging community; collaborative revision will be facilitated by the version control and source code management functionality provided by Git (and GitHub).

## 2. Data Levels for antibody-based multiplexed tissue Imaging

As of August 2021, the Minimal Information and Tissue Imaging standard (MITI) is still in an "request for comment" period, but we expect the final data levels for multiplexed tissue images generated using antibody reagents to be close to what is described below.

The concept of "data levels" (or tiers) was first developed by the Cancer Genome Atlas (TCGA), Genomic Data Commons (GDC) and the database of Genotypes and Phenotypes (dbGAP) to standardize the transformation of raw data (as generated by a measurement apparatus) into processed and interpreted data as used in research publications. Use of data levels promotes uniform and reproducible data analysis and interpretation [33]. Data levels for genomic data also distinguish between restricted access data that carries a de-identification risk (Level 1 and 2 data such as primary sequence) and open access data (Level 3 data such as RNA-seq gene counts). In the case of tissue images de-identification is not considered a risk and only the extent of data processing is considered in establishing a data level.

**Level 1 MITI data** comprise the raw numerical output of acquisition instruments (microscopes, slide scanner, mass cytometers etc.). These data may be in a variety of vendor-specific formats, although all microscope vendors and investigators are strongly encouraged to conform to universally recognized *Bioformats* standards. For a whole slide image, Level 1 will contain many individual image tiles (images recorded from different x, y positions in the specimen).

FASTq files are a common type of Level 1 data in genomics.

**Level 2 MITI data** comprise full-resolution primary images in the universal OME-TIFF format that have undergone stitching, registration, illumination correction, background subtraction, intensity normalization etc. to generate high quality mosaic images. The processing of Level 1 data to generate Level 2 data must be performed using automated software routines (not human intervention), ideally using open-source algorithms whose operation is transparent. The generation of an image mosaic from multiple image tiles using e.g., ASHLAR[43] is a prototypical Level 1 to Level 2 transformation.

BAM files are a common type of Level 2 data in genomics.

**Level 3 MITI data** are the results of image processing and include segmentation masks, images labelled by humans (e.g., to identify nuclei or annotated histology) or by software algorithms. The generation of Level 3 data may involve human interpretation, which should be recorded as part of the image metadata. To provide conformity with GDC data access concepts, full-resolution primary OME-TIFF images that have been subjected to human, or human-assisted software-based quality control are considered Level 3 data. A typical transformation from Level 2 to 3 images involves removing channels in which staining failed (a bad reagent batch) or cyclic data acquisition was interrupted (e.g., a dropped slide).

We anticipate that Level 3 images will be the primary types of images made available via public-facing data repositories. A key feature of microscopy in general, and tissue imaging in particular, is that virtually all types of human and computational analysis require access to full resolution data files, which can be very large. In contrast, in genomics, many types of analysis are possible using highly processed and compressed files; sustaining access to Level 3 images therefore imposes a substantial burden.

For tissue imaging, level 3 data types include:

- Level 3 Image mosaics that have been subjected to quality control, typically to remove staining and channel specific issues, and for cyclic methods, channels in which tissue damage has reached unacceptable levels.
- Segmentation masks, which are typically generated using software, such as UnMICST[44] but subjected to some level of human oversight or, in the case of machine learning, to supervised training. The models used to generate masks should be recorded.
- "Data Overviews" involving browser-based tools such as MINERVA[21] (or OME Viewers if an OMERO[25] database is available) that make it possible to browse images without having to download them. Viewing in these cases should involve as little additional interpretation as possible.

mRNA expression levels (RNA-seq gene count tables) are a common type of Level 3 genomic data.

**Level 4** data are numerical data generated from processing Level 3 data, most commonly to create "spatial feature tables" describing marker intensities, cell coordinates and other single-cell features (the analogy is with count tables in RNA sequencing).

**Level 5** data are results (e.g., cell type annotation) derived from Levels 4 spatial feature tables and level 3 images. Typical level 5 data include:

- MINERVA "Data Explorations" that use digital docents and human-generated annotation to guide users through the features of a complex set of images. The analogy is with a traditional figure.
- Dimensionality-reduced version of Level 4 data including all model parameters
- Machine learned models (other than segmentation models) from images or other numerical data
- Models that integrate image data with other data modalities
- Cell type and state annotations
- Tissue architecture information such as ducts in normal tissue and tumor nests in malignant tissue

### 3. Clinical and Patient-Derived Metadata

Clinical Data Elements are used to report patient related metadata including demographics, diagnosis, exposures, treatment and follow-up and are based on the Genomic Data Commons (GDC) Data Model. 'NOS' ('Not otherwise specified') indicates that a general diagnosis was possible but sufficient information was not available to provide a specific diagnosis. Complete data standards, including valid values, for HTAN are available on the HTAN Data Portal at https://htan-portal-nextjs.now.sh/standards. Note that metadata describing patient-derived material used to create a model may also be reported using these data elements where applicable.

**Supplementary Table 5. Clinical and Patient-Derived Data Elements for Highly Multiplexed Tissue Imaging Experiments**

| Attribute | Description | Valid Values | Significance |
|---|---|---|---|
| **Participant ID** | Participant Identifier | | REQUIRED |
| **Species** | Text that identifies species of tissue sample. | Human, Primate, Mouse, Other | REQUIRED |
| Demographics | | | |

| **Ethnicity** | Text designations that identify ethnicity. | Hispanic or Latino, Not Hispanic or Latino | REQUIRED |
|---|---|---|---|
| **Gender** | Text designations that identify gender. | Female, Male, non-binary, self-identify/other: (please specify) | REQUIRED |
| **Race** | Text designations that identify race. | White, American Indian or Alaska native, Black or African American, Asian, Native Hawaiian or other Pacific islander, other | REQUIRED |
| **Vital status at last follow up** | The survival state of the person at the last follow up contact. | Alive, Dead | REQUIRED |
| **Year of death** | Numeric value to represent the year of the death of an individual. | Integer (YYYY) | RECOMMENDED |
| **Cause of death** | Text term to identify the cause of death for a patient. | Cancer Related, Cardiovascular Disorder, NOS, End-stage Renal Disease, Infection, Not Cancer Related, Renal Disorder, NOS, Spinal Muscular Atrophy, Surgical Complications, Toxicity | RECOMMENDED |
| **Days to death** | Age at the time of death expressed in the number of days since birth. | Number | RECOMMENDED |
| Diagnosis ||||
| **Age at diagnosis** | Age at the time of diagnosis expressed in the number of days since birth. | Number | REQUIRED |
| **Known genetic predisposition mutation** | A yes/no/unknown indicator to identify whether there is a known genetic predisposition mutation present in the patient. | Yes, No | RECOMMENDED, IF APPLICABLE |
| **Hereditary cancer predisposition syndrome** | Inherited genetic predisposition syndrome that confers heightened susceptibility to cancer in the patient. | Examples for Oncology are listed in the NCI CDE Browser | RECOMMENDED, IF APPLICABLE |
| **Primary diagnosis** | Text term used to describe the patient's histologic diagnosis. | Examples for Oncology are listed in the NCI CDE Browser | REQUIRED |
| **Morphology** | Histologic Morphology Code based on International Classification of Diseases 10th Edition. | Examples for Oncology are listed in the NCI CDE Browser | REQUIRED |
| **Site of resection or biopsy** | The text term used to describe the anatomic site of origin, of the patient's disease. | Examples for Oncology are listed in the NCI CDE Browser | REQUIRED |
| **Tissue or organ of origin** | The text term used to describe the anatomic site of origin, of the patient's disease. | Examples for Oncology are listed in the NCI CDE Browser | REQUIRED |

| **Classification of tumor** | Text that describes the kind of disease present in the tumor specimen as related to a specific time point. | Primary, Metastasis, Recurrence, Progression, Premalignant | RECOMMENDED, IF APPLICABLE |
|---|---|---|---|
| **Prior treatment** | Yes/no/unknown indicator related to the administration of therapeutic agents received before the body specimen was collected. | Yes, No | RECOMMENDED, IF APPLICABLE |
| **Days to last follow up** | Time interval from the date of last follow up to the date of initial pathologic diagnosis, represented as a calculated number of days. | Number | REQUIRED |
| **Last known disease status** | Text term that describes the last known state or condition of an individual's neoplasm. | Distant met recurrence/progression, Loco-regional recurrence/progression, Biochemical evidence of disease without structural correlate, Tumor free, Unknown tumor status, With tumor | REQUIRED, IF APPLICABLE |
| **Days to last known disease status** | Time interval from the date of last follow up to the date of initial pathologic diagnosis, represented as a calculated number of days. | Number | REQUIRED, IF APPLICABLE |
| **Progression or recurrence** | Yes/No/Unknown indicator to identify whether a patient has had a new tumor event after initial treatment. | Yes, No | REQUIRED, IF APPLICABLE |
| **Days to progression** | Number of days between the date of the patient's first diagnosis and the date the patient's disease progressed. | Number | REQUIRED, IF APPLICABLE |
| **Days to progression free** | Number of days between the date used for index (DOB) and the date the patient's disease was formally confirmed as progression-free. | Number | REQUIRED, IF APPLICABLE |
| **Tumor grade** | Numeric value to express the degree of abnormality of cancer cells, a measure of differentiation and aggressiveness. | G1, G2, G3, G4, GX, GB, High Grade, Low Grade | REQUIRED, IF APPLICABLE |
| Exposures | | | |
| **Pack years smoked** | Numeric computed value to represent lifetime tobacco exposure defined as number of cigarettes smoked per day x number of years. | Number | RECOMMENDED |
| **Years smoked** | Numeric value to represent the number of years a person has been smoking. | Number | RECOMMENDED |
| Canonical mutational information | | | |
| **Gene symbol** | The text term used to describe a gene targeted or included in molecular analysis. | Refer to HTAN Data Portal for examples and complete list | REQUIRED, IF AVAILABLE |

| Molecular analysis method | The text term used to describe the method used for molecular analysis. | Refer to HTAN Data Portal for examples and complete list | REQUIRED, IF AVAILABLE |
|---|---|---|---|
| Test result | The text term used to describe the result of the molecular test. | Refer to HTAN Data Portal for examples and complete list | REQUIRED, IF AVAILABLE |
| Treatment | | | |
| Treatment type | Text term that describes the kind of treatment administered. | Examples for Oncology are listed in the NCI CDE Browser | REQUIRED |
| Initial disease status | Text term used to describe the status of the patient's malignancy when the treatment began. | Initial Diagnosis, Progressive Disease, Recurrent Disease, Residual Disease | RECOMMENDED, IF APPLICABLE |
| Therapeutic agents | Text identification of the individual agents used as part of a treatment regimen. | String | REQUIRED |
| Follow-Up | | | |
| Days to follow up | Number of days between the date of sample acquisition and the date of the patient's last follow-up appointment or contact. | Number | RECOMMENDED |
| Lost to follow up | Yes/No/Unknown indicator to identify whether a patient was lost to follow up. | Yes, No | RECOMMENDED |

## 4. Non-Human Metadata Elements for Highly Multiplexed Tissue Imaging Experiments

Separate data elements report on the creation, quality assurance, and study of tissues from non-human model organisms for imaging experiments. With mouse tissues being a prominent example, Supplementary Table 6 aggregates the existing standards defined for mouse models by the Mouse Tumor Biology (MTB) Database[45,46] and the Patient-Derived Xenograft Minimum Information (PDX-MI) Standards[15], in accordance with guidelines set forth by the Alliance for Genome Resources[47].

**Supplementary Table 6. Mouse Metadata Elements for Multiplexed Tissue Imaging Experiments**

| Attribute | Significance |
|---|---|
| Submitter Patient ID | REQUIRED |
| Submitter Sample ID | REQUIRED |
| Disease tissue of origin | REQUIRED |
| Disease type | REQUIRED |
| Is sample from untreated patient? | REQUIRED |
| Original sample type | RECOMMENDED |
| Sample from an existing PDX model | RECOMMENDED |
| Submitter PDX ID | REQUIRED |
| Mouse strain (and source) | REQUIRED |

| Attribute | Significance |
|---|---|
| Strain immune system humanized? | REQUIRED |
| Type of humanization | REQUIRED |
| Sample preparation | REQUIRED |
| Injection type and site | REQUIRED |
| Mouse treatment for engraftment | RECOMMENDED |
| Engraftment rate | RECOMMENDED |
| Engraftment time | RECOMMENDED |
| Sample characterization technology | REQUIRED |
| Sample confirmed not to be of mouse/EBV origin | REQUIRED |
| Response to standard of care (pharmacological positive control) | REQUIRED |
| Animal health status | REQUIRED |
| Passage QA performance | REQUIRED |
| Treatment, passage | RECOMMENDED |
| Treatment protocol | RECOMMENDED |
| Treatment response | RECOMMENDED |
| Sample OMICS | RECOMMENDED |
| Lag time/doubling time | RECOMMENDED |
| PDX model availability? | RECOMMENDED |

## 5. Biospecimen Metadata

For each participant in a study, one or several biospecimen samples may be analyzed. Each biospecimen should be assigned a unique label (Biospecimen ID). Relationships between participants and biospecimens can be indicated using the Parent ID; Parent IDs may be used to indicate the source of the biospecimen and when multiple biospecimens are derived from the same source. For tissue imaging, samples are either frozen or fixed and then sectioned and mounted onto slides. In this case each slide is a unique biospecimen from the same parent tissue block. An implementation example is included in Section 11 below.

These biospecimen metadata attributes include information about the method of tissue acquisition, sample processing and handling, and histologic assessment.

**Supplementary Table 7. Biospecimen Attributes for Highly Multiplexed Tissue Imaging Experiments**

| Attribute | Description | Valid Values | Significance |
|---|---|---|---|
| Biospecimen ID | Identifier for the Biospecimen | | REQUIRED |
| **Parent ID** | Parent Identifier from which the biospecimen was obtained. The parent could be another biospecimen or a research participant. | | REQUIRED |
| **Adjacent Biospecimen ID** | List of Identifiers (separated by commas) of adjacent biospecimens cut from the same sample | | REQUIRED |
| **Biospecimen Type** | Biospecimen Type | Tissue, Bone Marrow, Cell Block, Blood, Fluids | REQUIRED |
| **Analyte Type** | Text term that represents the kind of molecular specimen analyte. | Tissue section | REQUIRED |
| Collection and Processing | | | |
| **Protocol Link** | Identifier that describes the protocol by which the sample was obtained or generated. E.g. Protocols.io | A valid Digital Object Identifier (DOI) | REQUIRED |
| **Timepoint Label** | Label to identify the time point at which the biospecimen was obtained. | Baseline, End of Treatment, Overall survival, Final | REQUIRED |
| **Collection Days From Index** | Number of days from the research participant's index date that the biospecimen was obtained. | Number | REQUIRED |
| **Acquisition Method Type** | Records the method of acquisition or source for the specimen under consideration. | Autopsy, Biopsy, Fine needle aspirate, Surgical Resection, Punch biopsy, Shave biopsy, Excision, Re-excision, Sentinel node biopsy, Lymphadenectomy | REQUIRED |
| **Post-mortem Interval** | Number of days from the research participant's date of death that the biospecimen was obtained. | Number | RECOMMENDED |
| **Tumor tissue type** | Text that describes the kind of disease present in the tumor specimen as related to a specific time point. | Primary Tumor, Local Tumor Recurrence, Distant Tumor Recurrence, Metastatic, Premalignant | REQUIRED, IF APPLICABLE |
| **Preservation method** | Text term that represents the method used to preserve the sample. | Formaldehyde fixed paraffin embedded (FFPE), Frozen | REQUIRED |
| **Fixative Type** | The field to identify the type of fixative used to preserve a tissue specimen | Formaldehyde | REQUIRED |

| Fixation Duration | The length of time, from beginning to end, required to process or preserve biospecimens in fixative (measured in minutes) | Number | REQUIRED |
|---|---|---|---|
| Slide Charge Type | A description of the charge on the glass slide. | Uncharged, Charged, Coverslip, Coated, Not applicable | REQUIRED |
| Section Thickness Value | Numeric value to describe the thickness of a slice to tissue taken from a biospecimen, measured in microns (um). | Number | REQUIRED |
| Days to sectioning | Number of days between the date of sample acquisition and the date that the specimen was sectioned | Number | REQUIRED |
| Storage Method | The method by which a biomaterial was stored after preservation or before another protocol was used. | Ambient temperature, 4°C, -20°C, -80°C | REQUIRED |
| Days to processing | Number of days between the date of sample acquisition and the date that the biospecimen was processed | Number | REQUIRED |
| Shipping Conditions | Text descriptor of the shipping environment of a biospecimen. | Ambient Pack, Cold Pack, Dry Ice, Ice Pack, Liquid Nitrogen, Other Shipping Environment, Specimen at Room Temperature, Not shipped | REQUIRED |
| Histologic Assessment ||||
| Histology Assessments By | Text term describing who made the histological assessments of the sample | Pathologist, Research Scientist, Other, Unknown | REQUIRED |
| Histology Assessment Medium | The method of assessment used to characterize histology | Digital, Microscope | RECOMMENDED |
| Tumor infiltrating lymphocytes | Fraction of tumor-infiltrating lymphocytes | Number | RECOMMENDED, IF APPLICBALE |
| Degree of dysplasia | Information related to the presence of cells that look abnormal under a microscope but are not cancer. Records the degree of dysplasia for the cyst or lesion under consideration. | Normal, basal cell hyperplasia, or metaplasia Mild dysplasia, Moderate dysplasia, Severe dysplasia, Carcinoma in Situ | RECOMMENDED, IF APPLICABLE |
| Dysplasia fraction | Information related to the presence of cells that look abnormal under a microscope but are not cancer. Records the degree of dysplasia for | Number | RECOMMENDED, IF APPLICABLE |
| | the cyst or lesion under consideration. | | |
| **Number proliferating cells** | Numeric value that represents the fraction of proliferating cells determined during pathologic review of the sample slide(s). | Number | RECOMMENDED, IF APPLICABLE |
| **Percent Necrosis** | Numeric value to represent the percentage of cell death in a malignant tumor sample or specimen. | Number | RECOMMENDED, IF APPLICABLE |
| **Percent Normal Cells** | Numeric value to represent the percentage of normal cell content in a malignant tumor sample or specimen. | Number | RECOMMENDED, IF APPLICABLE |
| **Percent Stromal Cells** | Numeric value to represent the percentage of reactive cells that are present in a malignant tumor sample or specimen but are not malignant such as fibroblasts, vascular structures, etc. | Number | RECOMMENDED, IF APPLICABLE |
| **Percent Tumor Cells** | Numeric value to represent the percentage of infiltration by tumor cells in a sample. | Number | RECOMMENDED, IF APPLICABLE |
| **Percent Tumor Nuclei** | Numeric value to represent the percentage of tumor nuclei in a malignant neoplasm sample or specimen. | Number | RECOMMENDED, IF APPLICABLE |
| Imaging | | | |
| **Fiducial marker** | Fiducial markers for the alignment of images taken across multiple rounds of imaging. | Nuclear Stain (DAPI), Fluorescent beads, Grid slides (hemocytometer), Adhesive markers | REQUIRED (unless N/A) |
| **Mounting medium** | The solution in which the specimen is embedded, generally under a cover glass. It may be liquid, gum or resinous, soluble in water, alcohol or other solvents and be sealed from the external atmosphere by non-soluble ringing media. | Aqueous (water based) medium, Non-Aqueous (Solvent) based, Xylene, Toluene, Antifade with DAPI, Antifade without DAPI, PBS | RECOMMENDED |
| **Slicing method** | The method by which the tissue was sliced. | Vibratome, Cryosectioning, Tissue molds, Sliding microtome, Sectioning | RECOMMENDED |
| **Control** | Control tissue / area | Cell line, tissue, mutation, matrix, metal coating, fiducials | RECOMMENDED |

## 6. File-level Metadata (data level 1)

Raw files without any preprocessing steps are considered Level 1 and the following attributes are needed to sufficiently describe the individual files. These files are sometimes in proprietary formats. For some imaging assays such as whole slide scanning of H&E and IHC stained sections where the output of the machine is immediately usable and doesn't require any pre-processing (e.g., in Leica/Aperio-generated .svs-format files), a format translation may be the only step required to create the "Level 2" OME-TIFF file.

**Supplementary Table 8: File-level Metadata for Highly Multiplexed Tissue Imaging Experiments**

| Attribute | Description | Valid Values | Significance |
|---|---|---|---|
| Data File Id | unique e.g., "self" ID for data file described by this metadata | see HTAN identifier SOP or create a new one for your project | REQUIRED |
| Filename | name of data file as submitted to repository | no spaces, no special- characters | REQUIRED |
| File Format | format of data file | SVS, OME-TIFF, CZI | REQUIRED |
| Channel-Level Metadata CSV Filename | "companion" CSV file containing channel-level metadata for multi-channel/multi-cyclic methods | no spaces, no special- characters | REQUIRED (unless N/A) |
| Imaging modality | imaging technology | epifluorescence, lightsheet, confocal, ion beam, laser ablation | REQUIRED |
| Assay Type | method / technology used | H&E, t-CyCIF, IHC, mIHC, MxIF, SABER, IMC, CODEX, MIBI, MELC | REQUIRED |
| Protocol ID(s) | unique identifier (e.g. protocols.io) -- *may be a delimited **list** of identifiers* | DOI | REQUIRED |
| Acquisition software | method(s) and parameters used for any acquisition/ transformation/ processing to create this image | DOI or github repo/commit or name/version | RECOMMENDED |
| Image Acquisition Date | | YYYY-MM-DD | RECOMMENDED |
| Microscope | Microscope type (manufacturer, model, etc) used for this experiment | | RECOMMENDED |
| Objective | objective lens used (descriptor or other identifier) | | RECOMMENDED |
| Nominal Magnification | nominal optical magnification | 10X, 40X | RECOMMENDED |
| Lens NA | lens numerical aperture | (floating point) | RECOMMENDED |
| Working Distance | distance from objective to coverslip | (floating point) | RECOMMENDED |
| Working Distance Unit | default = microns | | RECOMMENDED |
| Immersion Medium | the imaging medium affects the working NA of the objective | Air, Water, Glycerin, Oil | RECOMMENDED |
| **Comment** | Free text field to use if necessary | | RECOMMENDED |
| **Frame_ Averaging** | Number of frames averaged together (if no averaging, set to 1) | 1, 2, 3, ... | RECOMMENDED |
| **Fov_Number** | Index of FOV (as it pertains to its place in the Experiment) | 1, 2, 3, ... | REQUIRED (unless N/A) |
| **Fov_Sizex** | FOV size in X-dimension (micron) | | REQUIRED (unless N/A) |
| **Fov_Sizey** | FOV size in Y-dimension (micron) | | REQUIRED (unless N/A) |
| **Pyramid** | whether or not image file contains a pyramid stack | True or False (or T/F) | REQUIRED (unless N/A) |
| **Z_Stack** | whether or not image file contains a Z-stack | True or False (or T/F, N/A) | REQUIRED (unless N/A) |
| **T_Series** | whether or not image file contains a time series | True or False (or T/F, N/A) | REQUIRED (unless N/A) |

## 7. Channel-level Metadata Attributes (data levels 1 and 2)

Channel-level metadata can be provided in a companion CSV file with each OME-TIFF file associated with one such table. This companion CSV, referenced by filename in the corresponding level 1 and 2 file level metadata table **(Supplementary Table 7),** should have as many rows as necessary to describe each channel in the image data file -- with at least one row of data for each channel within the image data file. Each column in the CSV will represent a single attribute defined below, for all channels. Antibodies must be identified with their RRID identifiers[2] and validation should follow previously described practices[1,48] although validation results are currently external to the MITI standard.

The first two attributes listed below are required in order to align other information in this CSV to the metadata in the OME-TIFF header and in the file-level metadata. Beyond these two, the submitting center has flexibility to choose what information to provide -- *but ideally the list of attributes should be finite and agreed-upon*. If there are channels which represent more than one target, or more than one antibody or fluorophore, then that information should be provided on multiple rows (and the corresponding Channel ID and Channel Name repeated) to follow tidy data standards[40].

**Supplementary Table 9. Channel-level Metadata for Highly Multiplexed Tissue Imaging Experiments**

| Attribute | Description | Valid Values | Significance |
|---|---|---|---|
| **Channel ID** | this must match the corresponding field in the OME-XML / TIFF header | "Channel:0:1" | REQUIRED |
| **Channel Name** | this must match the corresponding field in the OME-XML / TIFF header | "Blue" or "CD45" or "E-cadherin" | REQUIRED |
| **Cycle Number** | the cycle # in which the co-listed reagent(s) was(were) used | 1, 2, 3, … (up to number of cycles) | RECOMMENDED |
| **Sub-Cycle #** | sub-cycle # | 1, 2, 3, ... | RECOMMENDED |
| **Target name** | short descriptive name (abbreviation) for this target (antigen) | "Keratin", "CD163", "DNA" | REQUIRED |
| **Antibody Name** | short descriptive name for this antibody | "Keratin-570", "CD8a-488" | REQUIRED |

| Antibody Role | where appropriate, indicate whether antibody is primary or secondary | "Primary" or "Secondary" | RECOMMENDED |
|---|---|---|---|
| RRID identifier | Research Resource Identifier | "RRID: AB_394606" | REQUIRED |
| Fluorophore | Fluorescent dye label | "Alexa Fluor 488" | REQUIRED |
| Clone | Unique clone identifier | "OX-8" | REQUIRED |
| Lot | lot number from vendor | | REQUIRED |
| Vendor | Vendor name | Abcam, CST, eBioscience | REQUIRED |
| Catalog # | catalog number from vendor | | REQUIRED |
| Excitation wavelength | center/peak of the excitation spectrum (nm) | 499 | RECOMMENDED |
| Emission wavelength | center/peak of the emission spectrum (nm) | 520 | RECOMMENDED |
| Excitation bandwidth | nominal width of excitation spectrum (nm) | 30 | RECOMMENDED |
| Emission bandwidth | nominal width of emission spectrum (nm) | 30 | RECOMMENDED |
| Metal isotope: Element | Element abbreviation | "La" or "Nd" | REQUIRED |
| Metal isotope: Mass | Element mass number | "139" or "142" | REQUIRED |
| Oligo barcode: upper strand | DNA barcode used for labeling | "AATGGTAC" | REQUIRED |
| Oligo barcode: lower strand | DNA barcode used for labeling | "AATGGTAC" | REQUIRED |
| Dilution | Final dilution ratio used in experiment | 1:1000 | RECOMMENDED |
| Concentration | Final concentration used in experiment | 10 ug/mL | RECOMMENDED |
| Passed QC | Identify stains that did not pass QC but are included in the dataset. | T/F | RECOMMENDED |
| QC details | Comment on why QC failed | text | RECOMMENDED |
| Comment | Free text field to use if necessary | | RECOMMENDED |

## 8. OME-TIFF Header Metadata Attributes (data level 2)

OME-TIFF is an open file format that combines the TIFF format for storing binary pixel data with an OME-based XML metadata header. Using the Bio-Formats software plug-in proprietary file formats are converted into OME-based files that can be opened using any Bio-Formats compatible software.

**Supplementary Table 10. OME-TIFF Header Metadata for Highly Multiplexed Tissue Imaging Experiments**

| Attribute | Description | Valid Values | Significance |
|---|---|---|---|
| Image ID | unique internal Image identifier | "Image:0" | REQUIRED |
| Pixels BigEndian | boolean | true, false | REQUIRED |
| DimensionOrder | internal ordering of dimensions | "XYZCT" | REQUIRED |

| PhysicalSizeX | physical size of one pixel in x-dimension | "0.650" | REQUIRED |
|---|---|---|---|
| PhysicalSizeXUnit | unit for PhysicalSizeX | "µm" | REQUIRED |
| PhysicalSizeY | physical size of one pixel in y-dimension | "0.650" | REQUIRED |
| PhysicalSizeYUnit | unit for PhysicalSizeY | "µm" | REQUIRED |
| PhysicalSizeZ | physical size of one pixel in z-dimension | "0.650" | REQUIRED if size of Z > 1 |
| PhysicalSizeZUnit | unit for PhysicalSizeZ | "µm" | REQUIRED if size of Z > 1 |
| SizeC | number of channels | integer > 0 | REQUIRED |
| SizeT | number of time-points | integer > 0 | REQUIRED |
| SizeX | number of pixels in x-dimension | integer > 0 | REQUIRED |
| SizeY | number of pixels in y-dimension | integer > 0 | REQUIRED |
| SizeZ | number of z-planes | integer > 0 | REQUIRED |
| Type | bit depth for each pixel value | "uint16" or "float" | REQUIRED |
| PlaneCount | total number of planes in this Image | integer > 0 | REQUIRED |
| Channel Name | channel label for each channel in this image | "Blue" or "CD45" or "E-cadherin" | REQUIRED |
| Comment | Free text field to use if necessary | | RECOMMENDED |

## 9. Processing- and Segmentation level Metadata Attributes (data level 3)

Processing-level attributes describe steps taken to address issues at the pixel level that compromise downstream analyses. This may include novel steps to identify and correct for image artefacts and reject channels that do not meet data standards. Thus, the output file will be an OME-TIFF image of similar bit depth, lateral size and resolution to the level 2 OME-TIFF but may have fewer channels.

Segmentation-level attributes primarily include information on the generation of the individual masks (e.g., segmentation method, individual thresholds, cellular expansion etc.) and the represented object classes (e.g., Nuclei, Cytoplasm, Cell, Tumor, Stroma etc.). Thus, segmentation masks can represent either cellular compartments or larger cellular communities / tissue structures as well as the same classes segmented with different parameters / methods. This allows dozens of masks to be associated with individual images and image-derived features. Individual pixels in segmentation masks should be labelled with unique object IDs within the scope of each image mask. Background pixels are set to zero. All masks should be saved as TIFF files containing integer values (8/16/32bit). The size and resolution should correspond to the level 2 OME-TIFF image.

Intermediate steps during segmentation (probability maps, e.g., UNET) / overlapping mask from instance segmentation (e.g., MASK-R-CNN) should be converted to a single channel labeled mask as described above. Those files can be stored alongside level 3 metadata.

**Supplementary Table 11. Processed-data level Metadata for Highly Multiplexed Tissue Imaging Experiments**

| Attribute | Description | Valid Values | Significance |
|---|---|---|---|
| **Data Type** | Specify if OME-TIFF provided is a segmentation mask or QC-checked image | mask, image | REQUIRED |
| **Data File ID** | unique e.g., "self" ID for data file described by this metadata | see identifier SOP or create a new one for your project | REQUIRED |
| **Filename** | name of data file as submitted to repository | no spaces, no special-characters | REQUIRED |
| **File Format** | format of data file | OME-TIFF | REQUIRED |
| **Parent Data File ID** | ID of Data File from which this Data File was derived | see identifier SOP or create a new one for your project | REQUIRED |
| **Protocol ID(s)** | unique identifier (e.g., protocols.io) -- *may be a delimited **list** of identifiers* | DOI | REQUIRED |
| **Software and Version** | method(s) and parameters used for any quality control, transformation, and visualization to create this image | DOI or github repo/commit or name/version | REQUIRED |
| **Commit SHA** | Short SHA for software version | 8 hexadecimal characters (for github) | REQUIRED |
| **Passed QC** | did all masks pass QC | True or False (or T/F) | If 'Data type' == 'image', REQUIRED, otherwise RECOMMENDED |
| **Comment** | Free text field to use if necessary | | RECOMMENDED |
| **Object Class** | Defines if the mask delineates nucleus, the cytoplasm, plasma membrane, the whole cell or other | nucleus, cytoplasm, cell, plasma membrane, other | If `Data type` = `mask`, REQUIRED |
| **Object Class Description** | Free text description of object class | eg. "organelle" | If Object Class = ''other'', REQUIRED, otherwise RECOMMENDED |
| **Minimum Intensity Display Range** | Lower-bound intensity value | eg "100" | RECOMMENDED |
| **Maximum Intensity Display Range** | Upper-bound intensity value | eg "65000" | RECOMMENDED |
| **FOV_number** | Index of FOV (as it pertains to its place in the Experiment) | 1, 2, 3, … (where valid values start at ) | RECOMMENDED |
| **FOV_SizeX** | FOV size in X-dimension (micron) | | RECOMMENDED |
| **FOV_SizeY** | FOV size in Y-dimension (micron) | | RECOMMENDED |
| **Pyramid** | whether or not image file contains a pyramid stack | True or False (or T/F) | REQUIRED |
| **Z_stack** | whether or not image file contains a Z-stack | True or False (or T/F) | REQUIRED |
| **T_series** | whether or not image file contains a time series | True or False (or T/F) | REQUIRED |
| **PhysicalSizeX** | physical size of one pixel in x-dimension | eg "0.650" | REQUIRED |
| **PhysicalSizeXUnit** | unit for PhysicalSizeX | eg "µm" | REQUIRED |
| **PhysicalSizeY** | physical size of one pixel in y-dimension | eg "0.650" | REQUIRED |
| **PhysicalSizeYUnit** | unit for PhysicalSizeY | eg "µm" | REQUIRED |
| **PhysicalSizeZ** | physical size of one pixel in z-dimension | eg "0.200" | REQUIRED if size of Z > 1 |
| **PhysicalSizeZUnit** | unit for PhysicalSizeZ | eg "µm" | REQUIRED if size of Z > 1 |
| **Type** | bit depth for each pixel value | eg "uint16" | REQUIRED |

## 10. Object-level Metadata Attributes (data level 4)

Segmentation-level attributes are associated with a CSV, FCS, or H5AD, etc file containing single cell level measurements across all channels (see channel-level; **Supplementary Table 8**) that have successfully passed QC. Therefore, masks for single cells (see segmentation-level; **Supplementary Table 10**) that have also successfully passed QC are combined with the individual channels for quantification. Non-cellular level information can be added as a separate column into the CSV, FCS, or H5AD file and recorded in the cell-level attributes of the files. A Python package ANNData, used for handling similar information for single-cell omics[41], is particularly well-suited here.

**Supplementary Table 12. Cell-level Attributes for Highly Multiplexed Tissue Imaging Experiments**

| Attribute | Description | Valid Values | Significance |
|---|---|---|---|
| **Data File ID** | unique e.g., "self" ID for data file described by this metadata | see identifier SOP or create a new one for your project | REQUIRED |
| **Filename** | name of data file as submitted to repository | no spaces, no special- characters | REQUIRED |
| **File Format** | format of data file | CSV, FCS, h5ad | REQUIRED |
| **Parent ID** | Parent Identifier from which the biospecimen was obtained. The parent could be another | see identified SOP | REQUIRED |
| | biospecimen or a research participant. | | |
| Software and Version | method(s) and parameters used for any acquisition/ transformation/ processing to create this image | DOI for protocols.io or github repo/commit or name/version | REQUIRED |
| Commit SHA | Short SHA for software version | 8 hexadecimal characters (for github) | REQUIRED |
| Passed QC | did all images/channels pass QC | True or False (or T/F) | REQUIRED |
| Comment | Free text field to use if necessary | | RECOMMENDED |
| Header size | How many columns in the CSV / FCS | e.g. 120 | REQUIRED |
| Object classes included | Defines which cell compartment this mask pertains or which tissue structure | Nucleus, cytoplasm, whole cell, plasma membrane, other | REQUIRED |
| Object Class Description | Free text description of object class | eg. "organelle" | If Object Class = ''other'', REQUIRED, otherwise RECOMMENDED |
| FOV_number | Index of FOV (as it pertains to its place in the Experiment) | 1,2,3 (where valid values start at ) | RECOMMENDED |
| FOV_SizeX | FOV size in X-dimension (micron) | eg "100" | RECOMMENDED |
| FOV_SizeY | FOV size in Y-dimension (micron) | eg "100" | RECOMMENDED |
| Pyramid | whether or not image file contains a pyramid stack | True or False (or T/F) | REQUIRED |
| Z_stack | whether or not image file contains a Z-stack | True or False (or T/F) | REQUIRED |
| T_series | whether or not image file contains a time series | True or False (or T/F) | REQUIRED |
| Type | Bit depth of image | uint8, uint16, uint32 | REQUIRED |
| PhysicalSizeX | physical size of one pixel in x-dimension | eg "0.650" | REQUIRED |
| PhysicalSizeXUnit | unit for PhysicalSizeX | eg "µm" | REQUIRED |
| PhysicalSizeY | physical size of one pixel in y-dimension | eg "0.650" | REQUIRED |
| PhysicalSizeYUnit | unit for PhysicalSizeY | eg "µm" | REQUIRED |
| PhysicalSizeZ | physical size of one pixel in z-dimension | eg "0.200" | REQUIRED if size of Z > 1 |
| **PhysicalSizeZUnit** | unit for PhysicalSizeZ | eg "µm" | REQUIRED if size of Z > 1 |
| **Cell-state level** | Are cell types / states included? | Yes / No | REQUIRED |

## 11. Cell-state Level Metadata Attributes (data level 5)

Attributes associated with cell types and cell states can be specified as a separate Level 5 table or in additional columns that augment Level 4 information (**Supplementary Table 11**) to reduce storage size. Each cell can be associated with multiple types/states, but each association must be specified on a separate row in the corresponding level 5 data file. The cell type/state annotations are encoded by a set of keywords from a predefined dictionary cataloguing all possible states and allowing for "Other/Unknown". The dictionary itself is typically derived from known marker-cell type associations (e.g., databases, literature, etc.) or in a data-driven fashion via clustering and cluster annotations. In the metadata table, the dictionary is stored a semicolon-delimited list of keywords (**Supplementary Table 12**). We envision that future versions of the MITI standard will define cell type ontologies to capture hierarchical relationships between the various cell types and states (e.g., "T-cell" is a child node of "Lymphocyte", which is itself a child node of "Immune"), much like GO ontologies currently catalogue hierarchical associations between protein function terms[49]. When implemented, the dictionary field in the metadata table will be replaced by a reference to the cell type ontology resource and its specific version.

**Supplementary Table 13. Cell-state Attributes for Highly Multiplexed Tissue Imaging Experiments**

| Attribute | Description | Valid Values | Significance |
|---|---|---|---|
| **Data File ID** | unique e.g., "self" HTAN ID for data file described by this metadata | see HTAN identifier SOP or create a new one for your project | REQUIRED |
| **Filename** | name of data file as submitted to repository | no spaces, no special- characters | REQUIRED |
| **File Format** | format of data file | CSV or FCS | REQUIRED |
| **Parent ID** | Parent Identifier from which the biospecimen was obtained. The parent could be another biospecimen or a research participant. | | REQUIRED |
| **Protocol ID(s)** | unique identifier (e.g. protocols.io) -- *may be a delimited list of identifiers* | DOI | REQUIRED |
| **Software and Version** | method(s) and parameters used for any acquisition/ transformation/ processing to create this image; can be a reference to a workflow defined in a standard workflow language (e.g., CWL, Nextflow, Snakemake) | DOI or github repo/commit or name/version | REQUIRED |
| **Comment** | Free text field to use if necessary | | RECOMMENDED |
| **Header size** | How many columns do have in the CSV / FCS / hdf5 | 120 | REQUIRED |
| **Possible cell type / states format** | A semicolon-delimited list of possible keywords that each cell will | "tumor;immune;stromal;proliferating;quiesc | RECOMMENDED |

| | be annotated with | ent;...;other;unknown" | |
|---|---|---|---|
| **Cell type clustering / calling methods** | Methods used (connected to protocols.io) | PhenoGraph, k-means, etc. | RECOMMENDED |
| **Algorithm parameters** | Semicolon-delimited list and description of algorithm parameters | k=15;metric=Euclidean;... | RECOMMENDED |

## 12. Implementations

Example metadata is provided below for two different types of samples. First, a colorectal cancer specimen acquired from the Cooperative Human Tissue Network (CHTN) and used as part of an HTAN trans-network project (HTAN TNP CRC1). Second, a COVID-19 patient specimen. Clinical, Biospecimen and Imaging metadata (e.g., for t-CyCIF) are provided.

Full example implementation can be found here: https://docs.google.com/spreadsheets/d/1ZSSAxLJ1ci8XQqZch93VmTgZNY6zlt3lgvNiN4e_n-Y/edit?usp=sharing

### Example 1: Overview colorectal cancer example (CRC1)

| Attributes | Example values from HTAN HTAN TNP CRC1 |
|---|---|
| Participant ID | HTA13_1 |
| Gender | Male |
| Race | White |
| Vital status | Alive |
| Age at diagnosis | 25185 days |
| Primary diagnosis | Malignant adenocarcinoma |
| Morphology | Mixed mucinous and signet ring cell adenocarcinoma |
| Site of resection or biopsy | Cecum |
| Tissue or organ of origin | Cecum |
| Classification of tumor | Primary |
| Prior treatment | No |
| Days to last follow up | 720 |
| Last known disease status | Tumor free |
| Days to last known disease status | 720 |
| Progression or recurrence | Yes |
| Days to recurrence | 365 |
| Tumor grade | G3 |
| Gene symbol | Not applicable |
| Molecular analysis method | Microsatellite analysis |
| Test result | High |
| Gene symbol | BRAF |
| Molecular analysis method | Targeted sequencing |
| Test result | Positive |
| Pack years smoked | 10 |
| Years smoked | 10 |
| Treatment type | Chemotherapy |
| Initial disease status | Initial Diagnosis |
| Therapeutic agents | Oxaliplatin |
| Days to follow up | 25905 |
| Biospecimen Type | Tissue |
| Analyte type | Tissue section |
| Protocol link | dx.doi.org/10.17504/protocols.io.bji2kkge |
| Timepoint label | Initial diagnosis |
| Acquisition method type | Surgical Resection |
| Tumor tissue type | Primary Tumor |
| Preservation method | Formalin fixed paraffin embedded (FFPE) |
| Fixative type | Formalin |
| Fixation duration | 24 |
| Slide charge type | Charged |
| Section thickness value | 5 |
| Storage method | Refrigerated at 4 degrees |
| Shipping conditions | Ambient pack |
| Histology assessments by | Pathologist |
| Degree of dysplasia | Carcinoma in situ |
| Percent stromal cells | 40 |
| Percent tumor cells | 40 |
| Mounting medium | Xylene |
| Slicing method | Sectioning |

## Example 2: COVID-19 example

| Attributes | Example values from study of SARS-CoV-2 |
|---|---|
| Participant ID | S01 |
| Gender | Female |
| Race | Black/African American |
| Vital status | Dead |
| Age at diagnosis | 24090 days |
| Known genetic predisposition mutation | Not applicable |
| Hereditary cancer predisposition syndrome | Not applicable |
| Primary diagnosis | COVID-19 pneumonia |
| Morphology | U07.1 |
| Site of resection or biopsy | Lung, right upper lobe |
| Tissue or organ of origin | Lung, right upper lobe |
| Classification of tumor | Not applicable |
| Prior treatment | Not applicable |
| Days to last follow up | 21 |
| Last known disease status | Not applicable |
| Days to last known disease status | Not applicable |
| Progression or recurrence | Not applicable |
| Days to recurrence | Not applicable |
| Tumor grade | Not applicable |
| Pack years smoked | 0 |
| Years smoked | 0 |
| Initial disease status | Not applicable |
| Therapeutic agents | Azithromycin, Hydroxychloroquine, Tocilizumab |
| Days to follow up | 24111 |
| Biospecimen Type | Tissue |
| Analyte type | Tissue section |
| Protocol link | dx.doi.org/10.17504/protocols.io.bji2kkge |
| Timepoint label | Autopsy |
| Acquisition method type | Autopsy |
| Post-mortem interval | 2 |
| Tumor tissue type | Not applicable |
| Preservation method | Formalin fixed paraffin embedded (FFPE) |
| Fixative type | Formalin |
| Fixation duration | Not reported |
| Slide charge type | Charged |
| Section thickness value | 5 |
| Storage method | Refrigerated at 4 degrees |
| Shipping conditions | Not shipped |
| Histology assessments by | Pathologist |
| Degree of dysplasia | Not applicable |
| Percent stromal cells | Not applicable |
| Percent tumor cells | Not applicable |
| Mounting medium | Xylene |
| Slicing method | Sectioning |

## 13. Reporting Guidelines Discussion

These guidelines for reporting highly multiplexed tissue imaging experiments have been developed as part of discussions in the Human Tumor Atlas Network (HTAN) working groups (Clinical/Biospecimen Working Group, Molecular Characterization Working Group, Data Analysis Working Group, Policy Working Group), and have been modified through a series of HTAN-wide request for comments led by the HTAN Data Coordinating Center (DCC) and through discussions with members of the Image Analysis Working Group of the Cancer Systems Biology Consortium and Physical Sciences-Oncology Network (CSBC/PS-ON) communities, the National Cancer Institute Imaging Data Commons (IDC),

Human Cell Atlas (HCA), the Human BioMolecular Atlas Program (HuBMAP), cBioPORTAL for Cancer Genomics, and the Open Microscopy Environment (OMERO).